\documentclass[a4paper,11pt]{article}
\pdfoutput=1
\usepackage{jcappub} 

\allowdisplaybreaks

\usepackage{array}
\newcolumntype{P}[1]{>{\centering\arraybackslash}p{#1}}

\title{\boldmath Extracting cosmological parameters from N-body simulations using machine learning techniques}

\author{Andrei Lazanu}

\affiliation{Laboratoire de Physique de l’Ecole normale sup\'erieure, ENS, Universit\'e PSL, CNRS, Sorbonne Universit\'e, Universit\'e de Paris, F-75005 Paris, France}

\emailAdd{andrei.lazanu@ens.fr}

\abstract{We make use of snapshots taken from the \textsc{Quijote} suite of simulations, consisting of 2000 simulations where five cosmological parameters have been varied ($\Omega_m$, $\Omega_b$, $h$, $n_s$ and $\sigma_8$) in order to investigate the possibility of determining them using machine learning techniques. In particular, we show that convolutional neural networks can be employed to accurately extract $\Omega_m$ and $\sigma_8$ from the \textit{N}-body simulations, and that these parameters can also be found from the non-linear matter power spectrum obtained from the same suite of simulations using both random forest regressors and deep neural networks. We show that the power spectrum provides competitive results in terms of accuracy compared to using the simulations and that we can also estimate the scalar spectral index $n_s$ from the power spectrum, at a lower precision.}

\begin{document}
\maketitle
\flushbottom

\section{Introduction}
In recent years, cosmology has increasingly become a precision science and measurements of cosmological observables have achieved an unprecedented level of accuracy. Probes of the Cosmic Microwave Background (CMB), such as COBE, WMAP \cite{2003ApJ...583....1B} and \textit{Planck} \cite{Akrami:2018vks}, have been able to measure it with exquisite accuracy and to show that the inflationary paradigm, combined with the growth of primordial quantum fluctuations, provide an excellent agreement with the cosmological data: the six-parameter $\Lambda$CDM model \cite{RevModPhys.61.1}. These parameters, together with others that can be derived from them, have been determined increasingly accurately, the current values of the ones used in the \textit{Planck} analysis being: the physical baryon density parameter $\Omega_b h^2 = 0.02242 \pm 0.00014$, the physical dark matter density parameter $\Omega_c h^2 = 0.11933 \pm 0.00091$, the approximation to the acoustic scale angle $100 \,\theta_{\rm MC}=1.04101 \pm 0.00029$, the Thomson scattering optical depth due to reionization $\tau=0.0561 \pm 0.0071$, the power spectrum of curvature perturbations $\ln (10^{10} A_s)=3.047 \pm 0.014$, the scalar spectral index $n_s=0.9665 \pm 0.0038$ (with error bars at 68\% confidence level), when considering \textit{Planck} and Baryon Acoustic Oscillations data. Derived parameters include the reduced Hubble constant $h = 67.66 \pm 0.42$, the baryon density parameter $\Omega_b = 0.0486 \pm 0.0010$, the  matter density parameter $\Omega_m  = 0.3111 \pm 0.0056$ and the amplitude of the root-mean-square matter fluctuation
averaged over a sphere of radius  at $8h^{-1} \mathrm{Mpc}$, $\sigma_8 = 0.8102 \pm 0.0060$. 
The CMB represents two-dimensional data from the surface of last scattering, thus providing only a limited number of modes. Moreover, it has already been exploited almost to the limit of cosmic variance. At the same time, the distribution of matter and galaxies -- the large scale structure of the Universe (LSS) -- contains significantly more modes, due to its three-dimensional nature given by the redshift, in addition to the distribution of galaxies on the sky. The observation of LSS, the modelling and the understanding of its nature are some of the goals of future probes of the large scale structure of the Universe, such as DESI \cite{Levi:2013gra}, Euclid \cite{Laureijs:2011gra}, LSST \cite{Abell:2009aa} and SKA \cite{Jarvis:2015tqa}. They will be providing in the near future more and more precise measurements of the galaxy distribution. The information encoded in the LSS, complementary to that from the CMB, is however much more difficult to extract, partially due to non-linear mode coupling and its three-dimensional nature. One cannot rely solely on analytical models or perturbative techniques --  $N$-body simulations are generally required, which are usually expensive to run. 
In parallel, the advances in computational power and processor architecture have allowed the running of higher and higher resolution numerical simulations, which model the current Universe increasingly realistically. Moreover, this has also allowed storing the outputs of many of these simulations, corresponding to different realisations or to variation of cosmological parameters. Advances in machine learning and data science techniques have allowed the efficient extraction of information from huge datasets. These  have become increasingly popular as the amount of information extracted from cosmological surveys has become overwhelming. Machine learning techniques have already been used in a variety of cosmology setups: CMB \cite{Caldeira:2018ojb, Chanda:2021qbf}, LSS \cite{Rodriguez:2018mjb,Lucie-Smith:2020ris,Lin:2021dim,Xu:2021pdp}, reionization and 21cm \cite{Shimabukuro:2017jdh,Huang:2020wvt}, gravitational lensing: weak lensing \cite{Ribli:2018kwb,ZorrillaMatilla:2020doz}, strong lensing \cite{Jacobs:2017xhn,Park:2020eat}, redshift prediction \cite{Collister:2003cz,Eriksen:2020diu}, parameter estimation \cite{Alsing:2019xrx, Kostic:2021tyw}, and are expected to provide more insights in the future \cite{Ntampaka:2019udw}.
In particular, such techniques can be applied to observations of the large-scale structure measured from galaxy surveys. 

In this work, we use machine learning methods to extract cosmological parameters from numerical simulations and from the non-linear power spectrum as a first step. We leave extensions to galaxy measurements to a future work. The paper is structured as follows: in Section \ref{sec:sims} we describe the simulations used in this work, in Section \ref{sec:fromsim} we show how one can extract the input parameters directly from the simulations, in Section \ref{sec:fromps} we use the non-linear power spectrum to extract the cosmological parameters and in Section \ref{sec:disc} we discuss the results. 

\section{Numerical simulations and model performance analysis}
\label{sec:sims}
In order to extract the cosmological parameters, we use  a subset of 2000 simulations of the \textsc{Quijote} simulations \cite{Quijote_sims}, a public suite  of 44100 full $N$-body simulations, ran using the TreePM code Gadget-III \cite{Springel:2005mi} in boxes of sides of $1\, {\rm Gpc}/h$. The authors provide a variety of cosmological results in addition to the snapshots of the simulations. In this work, we make use of the three-dimensional density field and the power spectrum measured from the simulations at redshift $z=0$. The simulations are run starting from $z=127$ and then evolved in time, where the matter power spectrum and the transfer functions are obtained from CAMB \cite{Lewis:1999bs} and suitably rescaled. These are used to determine  displacements and peculiar velocities using second order perturbation theory, which in turn are used to assign to particles that are initially laid on a regular grid with the 2LPT code \cite{Scoccimarro:1997gr, Crocce:2006ve}. The simulations have a cosmological volume of $1\, (h^{-1} {\rm Gpc})^3$. The gravitational softening length is set to 1/40 of the mean interparticle distance. The simulations have been run with five parameters  using  a latin-hypercube sampling, which is a statistical method for generating a near-random sample of parameter values from a multidimensional distribution \cite{10.2307/1268522}. The hypercube defining the parameter range is given by: $\Omega_m \in [0.1,0.5]$, $\Omega_b \in [0.03,0.07]$, $h \in [0.6,0.9]$, $n_s \in [0.8,1.2]$ and $\sigma_8 \in [0.6,1]$.

After splitting the data into training and test sets, we determine the cosmological parameters using the methods described in the next Sections. We then evaluate the performances of each model, presenting the results in two ways: (i) we plot the predicted value vs the ground truth for the test set; (ii) we determine the relative squared error (RSE) on the test set for each of the parameters in order to quantify the results. The RSE is given by
\begin{equation}
RSE = \frac{\sum_{i=1}^{n_{\rm test}} (y_{\rm true}^{(i)}-y_{\rm pred}^{(i)})^2}{\sum_{i=1}^{n_{\rm test}} (y_{\rm true}^{(i)}-\bar{y}_{\rm true})^2}    \,,
\end{equation}
where $n_{\rm test}$ represents the number of examples in the test set, $y_{\rm true}^{(i)}$ and $y_{\rm pred}^{(i)}$ represent the ground truth and predicted value of example $i$, and $\bar{y}_{\rm true}$ is the average of the ground truth values of the test set for the parameter in question. In the case of a good parameter determination we expect this quantity to be as close to 0 as possible.

\section{Cosmological parameters from the three-dimensional density field}
\label{sec:fromsim}
We use the three-dimensional distribution of the density field, interpolated on a $64^3$ grid from the standard resolution simulations ($512^3$ points) to extract the input cosmological parameters. As increasing the grid resolution significantly increases both the memory usage and the execution time, we restrict ourselves to this resolution in this work,  and we leave a more comprehensive analysis of the effects of the input size for future work. We note that this interpolation was readily available from the simulations website. We employ a set of 2000 simulations, where the five cosmological parameters have been varied: $\Omega_m$, $\Omega_b$, $h$, $n_s$ and $\sigma_8$. We are also in possession of the ``true'' values of the parameters used in the simulations. In order to test the performance of our model, we split the set consisting of 2000 simulations into a training set of 1600 simulations and a test set of 400 simulations.

\begin{figure}[!h]
\centering
\includegraphics[width=0.99\linewidth, trim=0 14cm 14cm 0, clip]{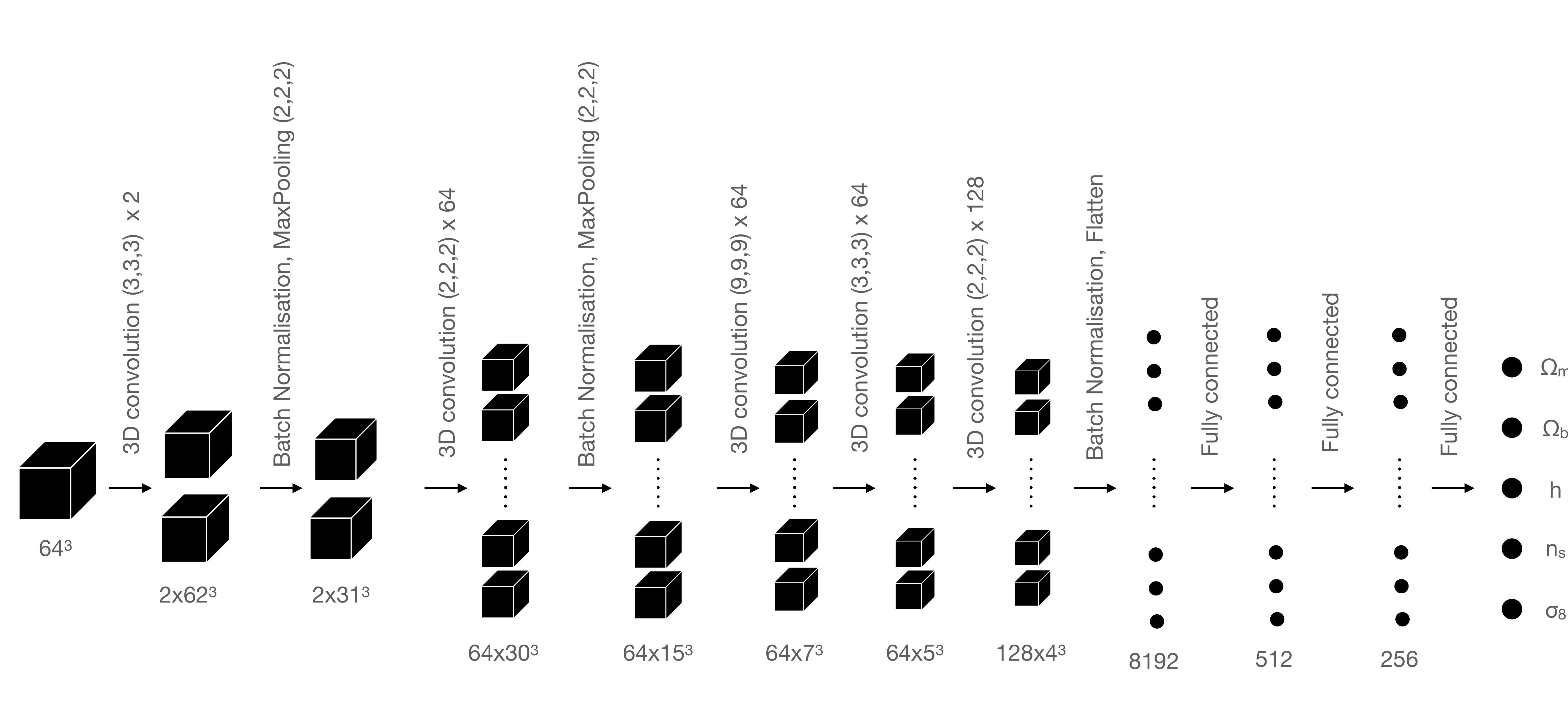} 
\caption{Architecture of the convolutional deep neural network. Starting with a cube of size $64^3$, the network consists of several convolutional layers, followed by Batch Normalisation layers to improve convergence and Max Pooling layers to reduce the space dimensionality. Finally, the network is flattened and contains two fully connected hidden layers, before the five-neuron output corresponding to the original input parameters $\Omega_m$, $\Omega_b$, $h$, $n_s$ and $\sigma_8$.}
\label{fig:cnn5}
\end{figure}

In order to take advantage of the three-dimensional nature of the data, we employ a deep 3D convolutional network. We start  with a convolutional neural network that aims to determine the five cosmological parameters. We have investigated several architectures, with the best one presented schematically in Fig. \ref{fig:cnn5}. 

\begin{figure}[!h]
\centering
\includegraphics[width=0.49\linewidth]{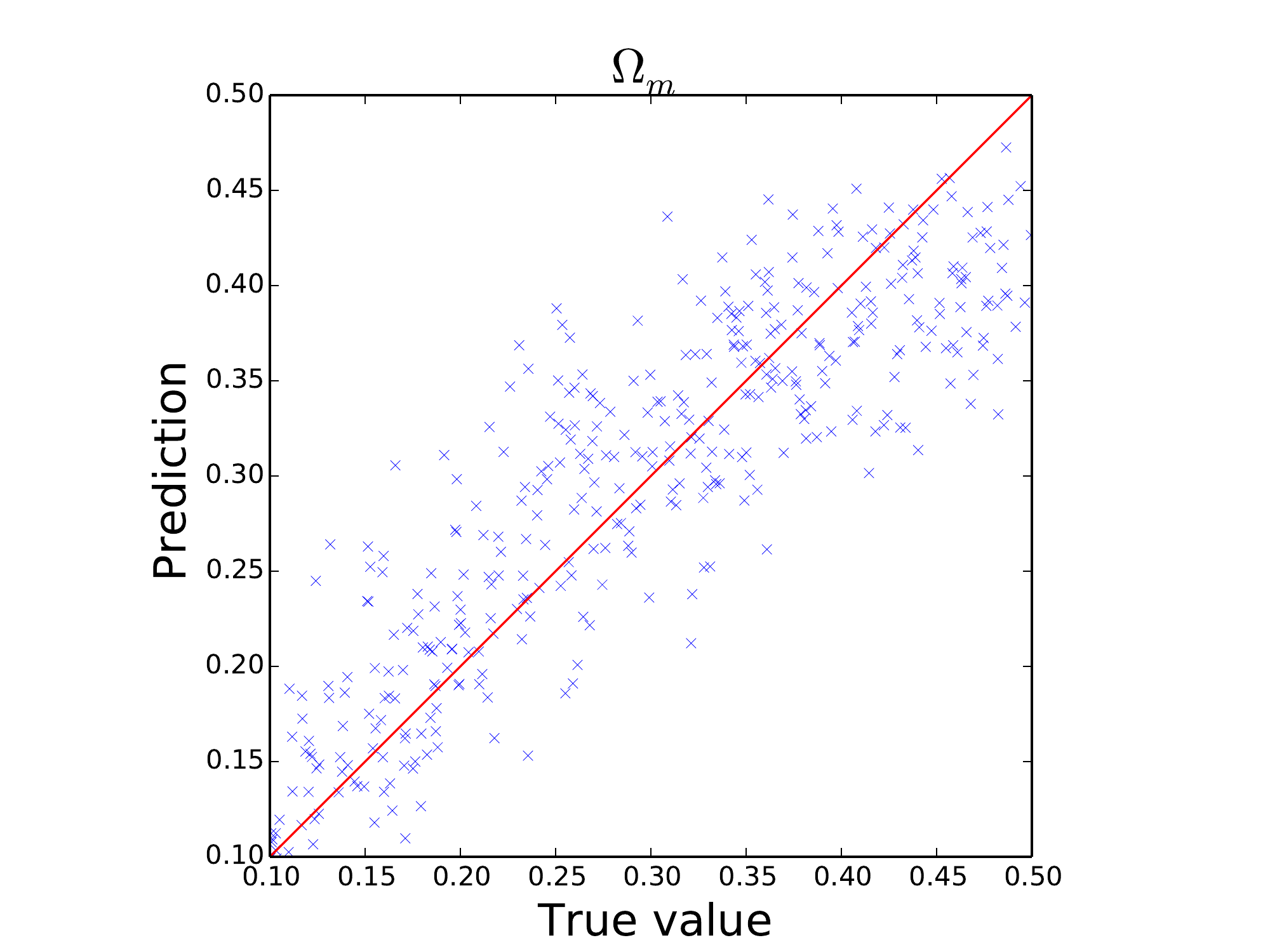} 
\includegraphics[width=0.49\linewidth]{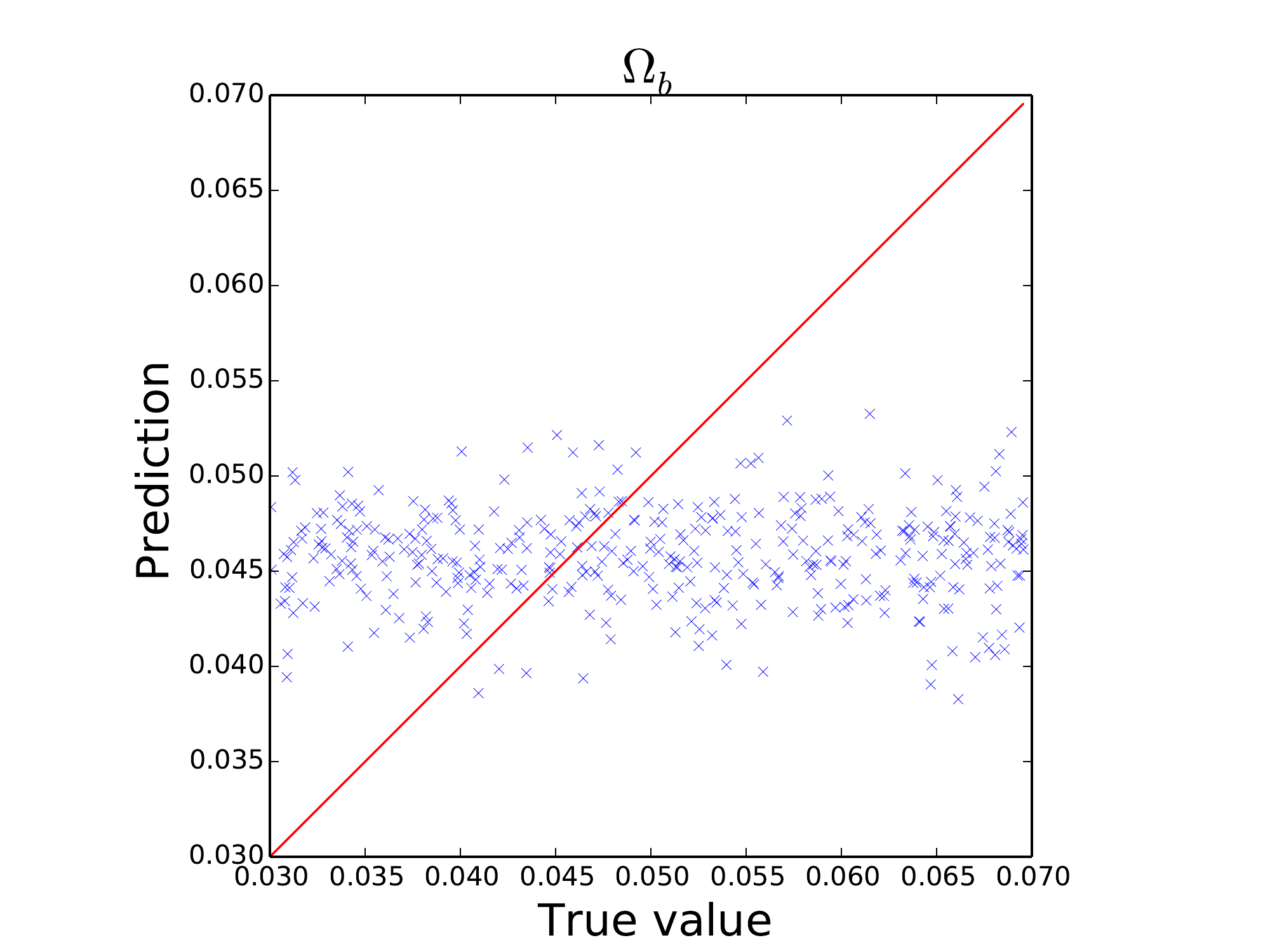} 
\includegraphics[width=0.49\linewidth]{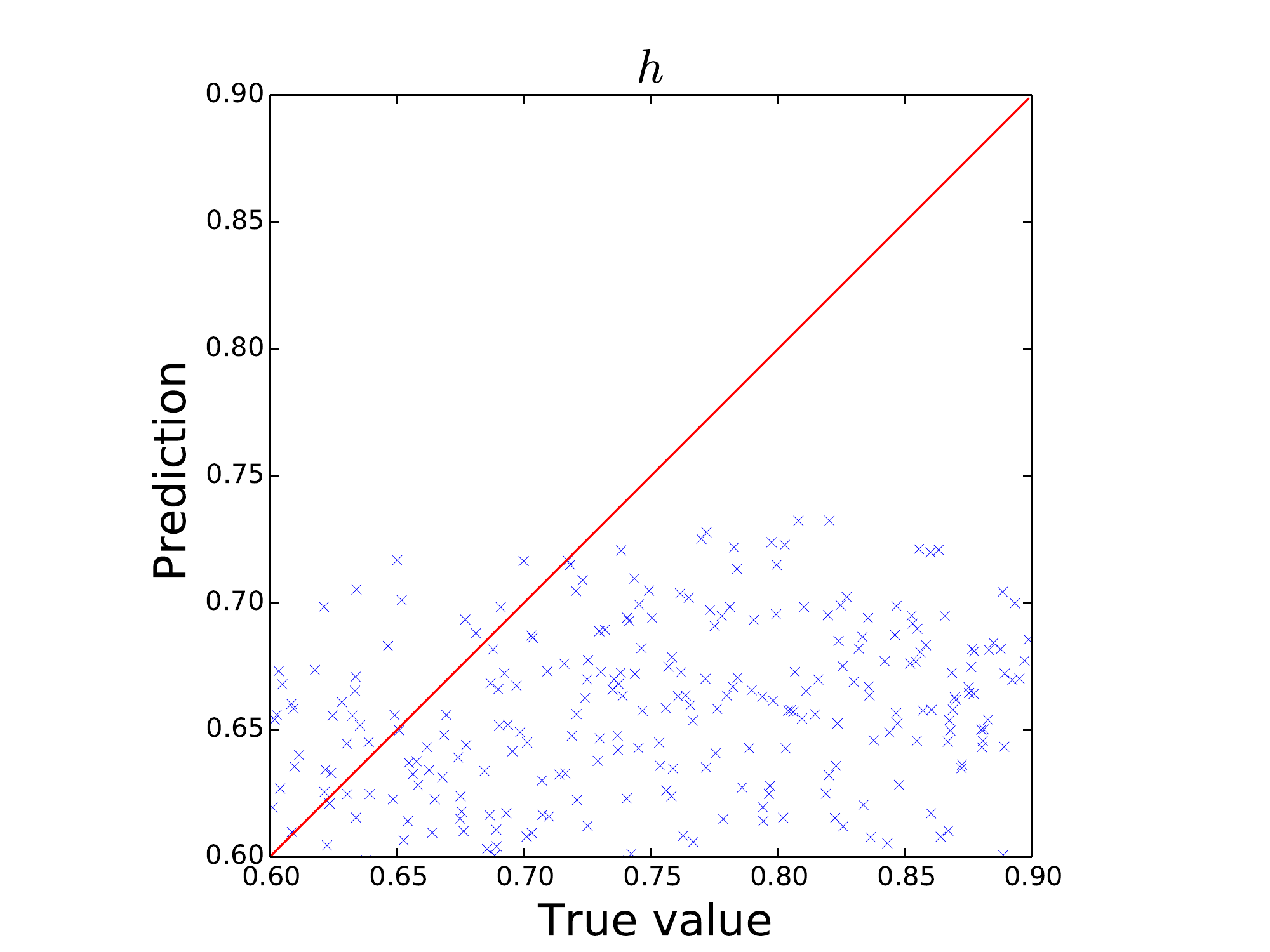} 
\includegraphics[width=0.49\linewidth]{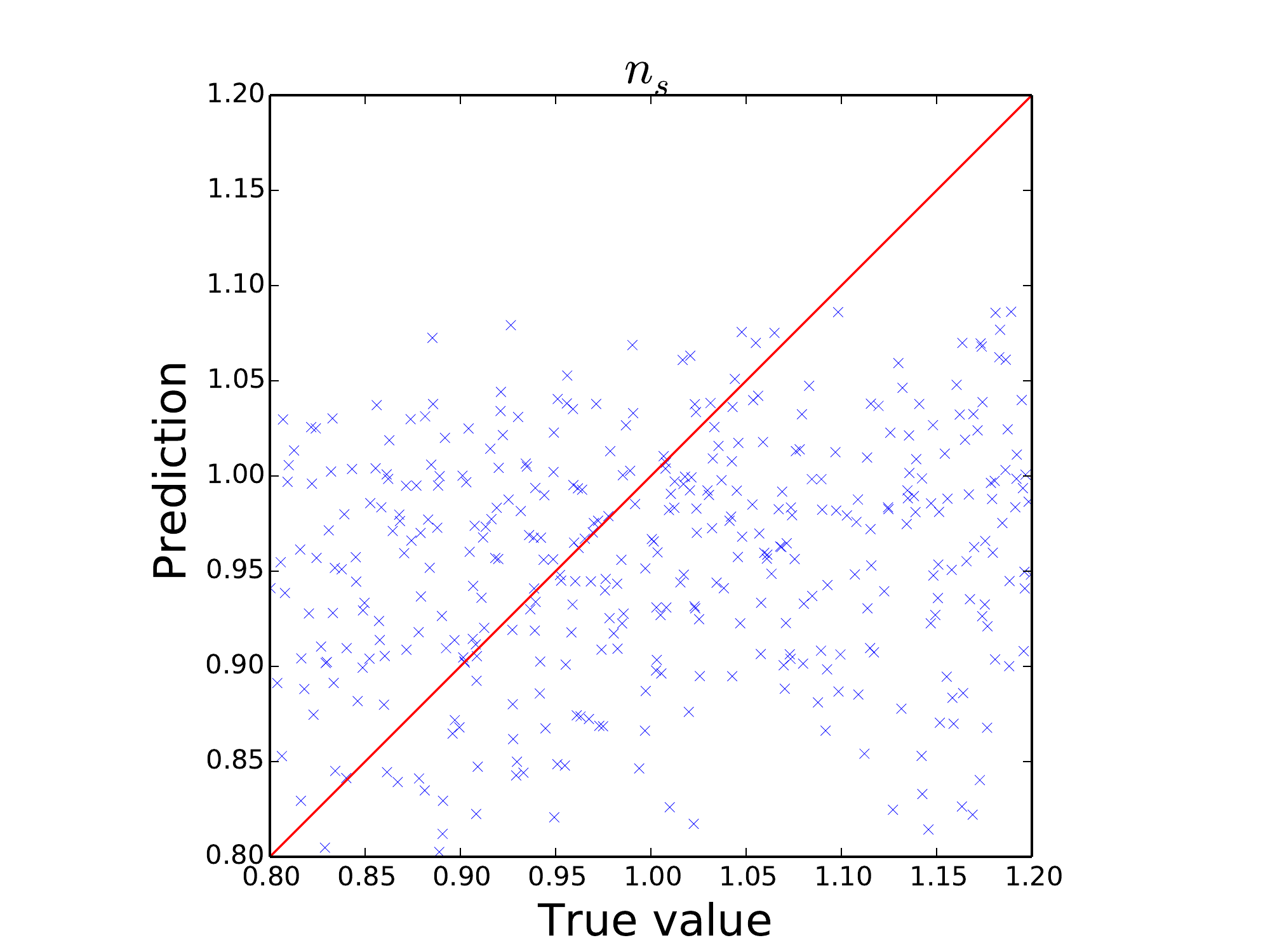} 
\includegraphics[width=0.49\linewidth]{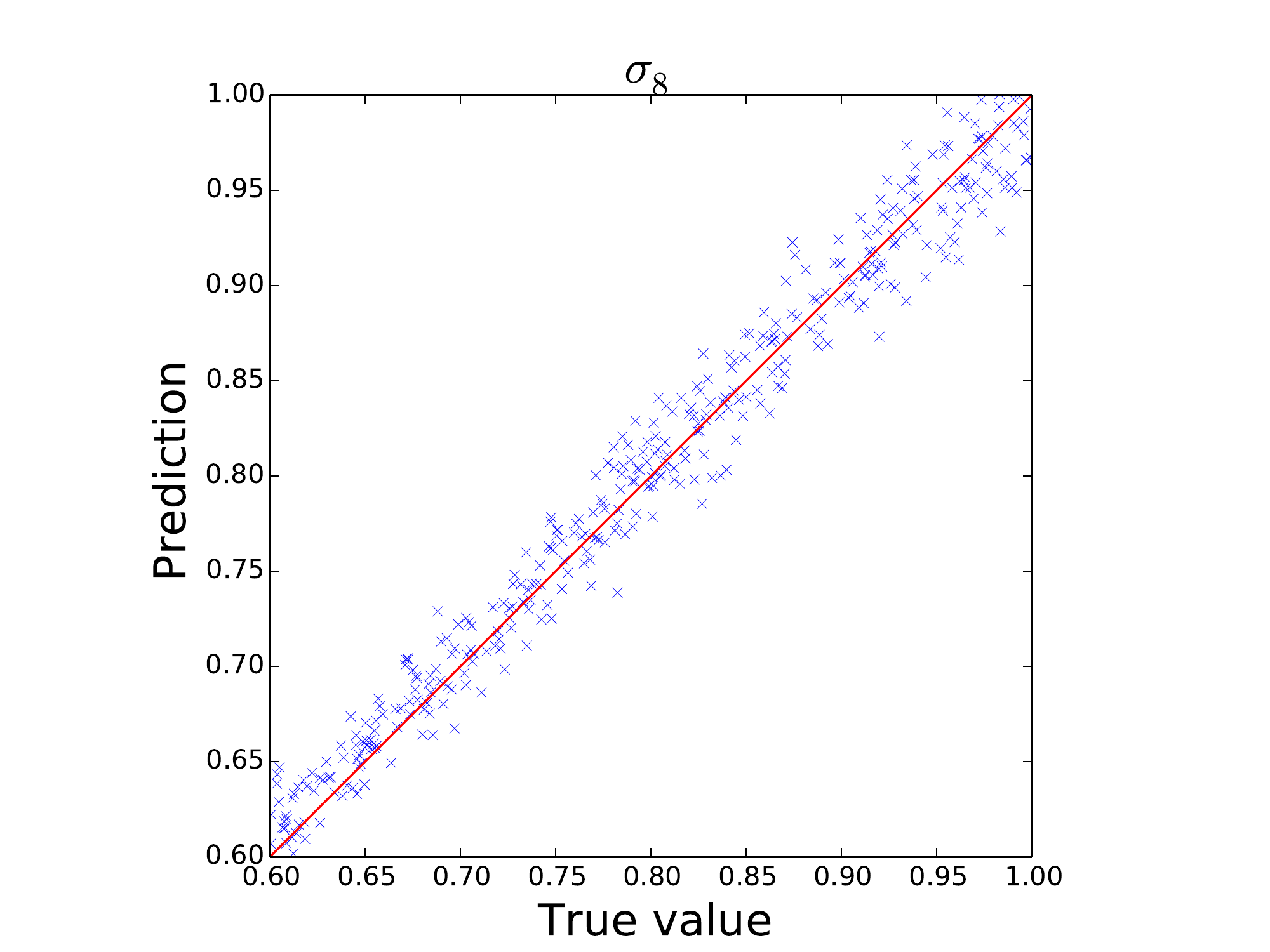} 
\caption{Predictions of $\Omega_m$, $\Omega_b$, $h$, $n_s$ and $\sigma_8$ (right) from the three-dimensional density field extracted from $N$-body simulations using a convolutional neural network.}
\label{fig:sims5}
\end{figure}

The network consists of three types of layers: 3D convolutions, batch normalisation and fully connected layers. We start with a $64^3$-voxel input layer, corresponding to the normalised density field and we consider two 3D convolutions, with a kernel size of $(2,2,2)$, followed by a batch normalisation and a max-pooling layer of size $(2,2,2)$ to reduce dimensionality. This is followed by 64 $(2,2,2)$ convolutions, a batch normalisation layer and a max-pooling layer of size $(2,2,2)$. We then add two 3D convolutional layers, of 64 channels and kernel sizes of $(9,9,9)$ and  $(3,3,3)$, followed by a 3D convolutional layer with 128 filters and a kernel of size $(2,2,2)$, followed by batch normalisation and flattening, now getting into a standard deep neural network, where we add two new hidden layers, with 512 and 256 neurons, before the five-neuron output layer, corresponding to the five parameters that have been varied in the simulations. We employ rectified linear unit (ReLU) activation functions throughout the network (defined as $f(x)=\max(0,x)$) and in order to regularise it we have chosen to use two L2 regularisers of size 0.05. We  use an Adam optimiser \cite{Kingma:2014vow} with a learning rate of $5 \times 10^{-5}$ and default first and second moment exponential decay rates of 0.9 and 0.999 respectively. For the metric, we look at the mean-squared error. 
We train the network for 50 epochs before it starts overfitting the training set.  We have investigated several choices for the regularisation techniques and for the values of its parameters, with the one presented providing the best results. The results on the test set are shown in Fig. \ref{fig:sims5} for each of the five parameters. We observe that $\sigma_8$ is the most accurately predicted, followed by $\Omega_m$, with RSEs of 0.025 and 0.22 respectively. Hence, our analysis shows that, out of the five parameters, the network was able to accurately predict $\Omega_m$ and $\sigma_8$, in line with what was found in Ref. \cite{Quijote_sims} from the probability density function.  This is most likely due to the weak dependence of the three-dimensional matter overdensity on the other three parameters.  We have investigated several architectures before choosing the one presented above. In particular, we have looked at the architecture that was used in Ref. \cite{Ravanbakhsh:2017bbi}, where, apart from the different number of kernels, the authors have used average-pooling layers instead of max-pooling used in our work and a Leaky rectified linear unit  \cite{Maas13rectifiernonlinearities} instead of the standard ReLU activation. These modifications didn't improve the results in our case.

We therefore concentrate on building a neural network that is able to accurately predict these two parameters. 
In this case we get to the convergence point after fitting the parameters on the training set for 30 epochs, getting RSEs of 0.23 and 0.015 respectively for $\Omega_m$ and $\sigma_8$ respectively.  
We note that the accuracy of the determination of $\sigma_8$ is improved, while for $\Omega_m$ we obtain a similar result as before. The better accuracy in the prediction of $\sigma_8$ using convolutional neural networks confirms the results from Ref. \cite{Ravanbakhsh:2017bbi}, where custom-run simulations have been employed, while another study of such parameter estimation technique can be found in Ref. \cite{Pan:2019vky}. 
As the accuracy of the determination of these two parameters is weaker with respect to that found in \cite{Pan:2019vky}, we have also investigated the possibility of splitting $128^3$ simulations into several $64^3$ ones in order to increase the training set size; however, the accuracy of the determination of the parameters was not improved.

\section{Cosmological parameters from the power spectrum}
\label{sec:fromps}
In this section, we describe how the power spectrum can be used to extract the cosmological parameters. We note that in Ref. \cite{Quijote_sims}  the probability density function has been used to extract the five parameters using a random forest regressor.

The power spectrum $P(k)$ represents the two-point correlation function (in Fourier space) of the matter overdensity,
\begin{equation}
\langle \delta(\mathbf{k}) \delta(\mathbf{k}')\rangle = (2 \pi)^3 \delta_D(\mathbf{k}+\mathbf{k}') P(k) \, ,
\end{equation}
where $\delta$ is the matter overdensity and $\delta_D$ is the Dirac delta function. In the case of Gaussian fields, the power spectrum contains all the information of the distribution \cite{Verde:2007wf}. For three-dimensional large-scale structure probes, the density fields are non-Gaussian on non-linear scales, and therefore contain additional information with respect to the linear power spectrum. Due to their non-linear nature,  it is not possible to determine precisely how much of the total information they encode, and how much is stored in higher-order statistics. 

Hence, we use the two-point function from the 2000 simulations and we try to extract the five cosmological parameters. As the scale in Fourier space is the same throughout the dataset, we can concentrate  only on the column corresponding to the values of the power spectrum. In this case, we use the  non-linear power spectrum extracted from the highest resolution \textsc{Quijote} simulations and we use random forest regressors \cite{Breiman_2001} as well as a deep neural network to extract the parameters. In order to improve the precision and the convergence speed of our computations, we take the logarithm of the power spectrum and in the case of the deep neural network we also normalise it.  This preprocessing step is justified by a decrease of the dynamical range of the input variable prior to any normalisation, which makes it more suitable for the use in a machine learning context. We note that $\sigma_8$ is determined as an integral over the power spectrum and $n_s \equiv \frac{d \ln P_{\rm lin}}{d \ln k}$;  therefore both can be obtained analytically given the power spectrum. Nevertheless, as these parameters are given as labels to the simulations and to the spectra, we proceed to determine them using random forests and neural networks.
\begin{figure}[!h]
\centering
\includegraphics[width=0.49\linewidth]{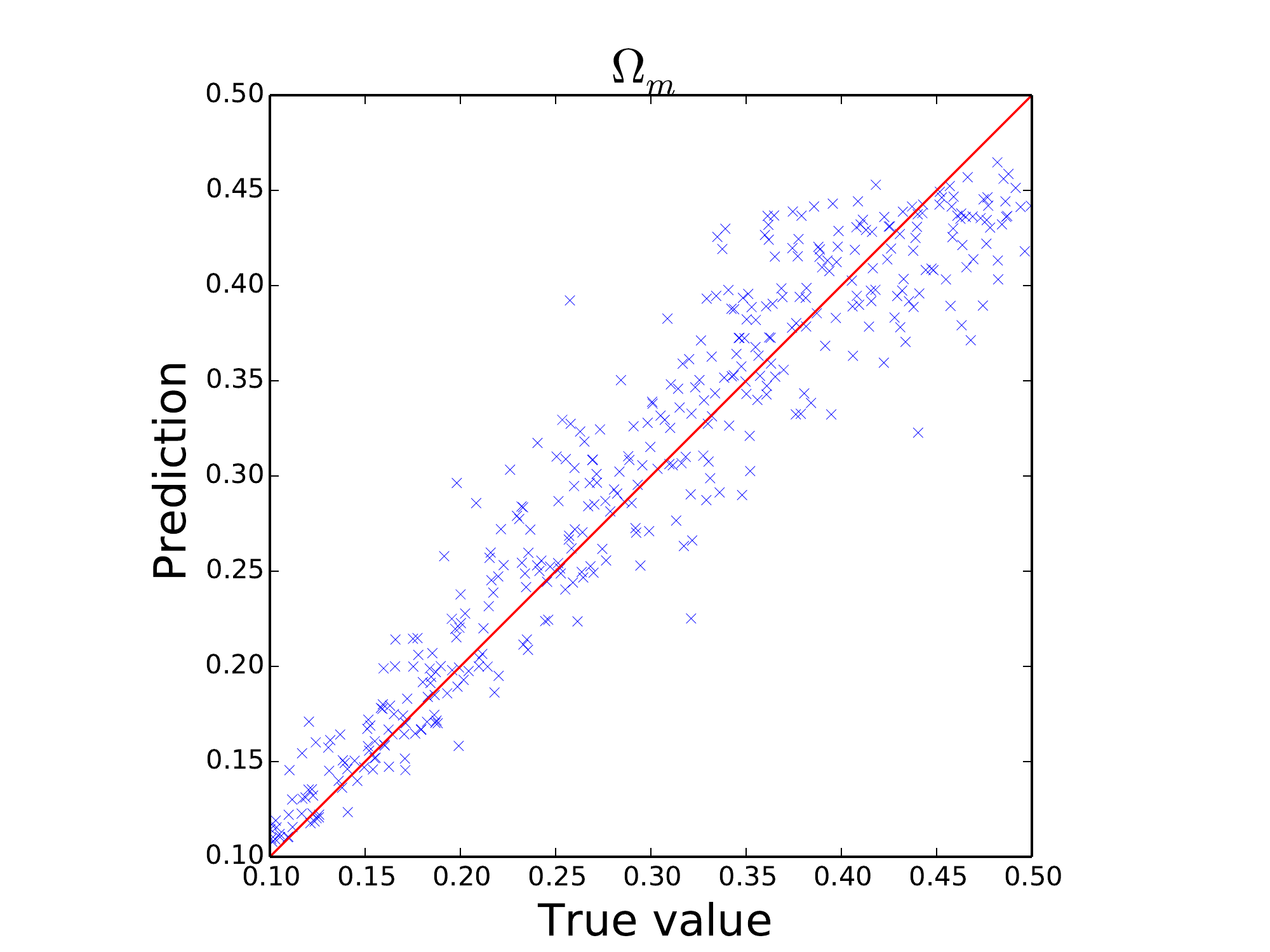} 
\includegraphics[width=0.49\linewidth]{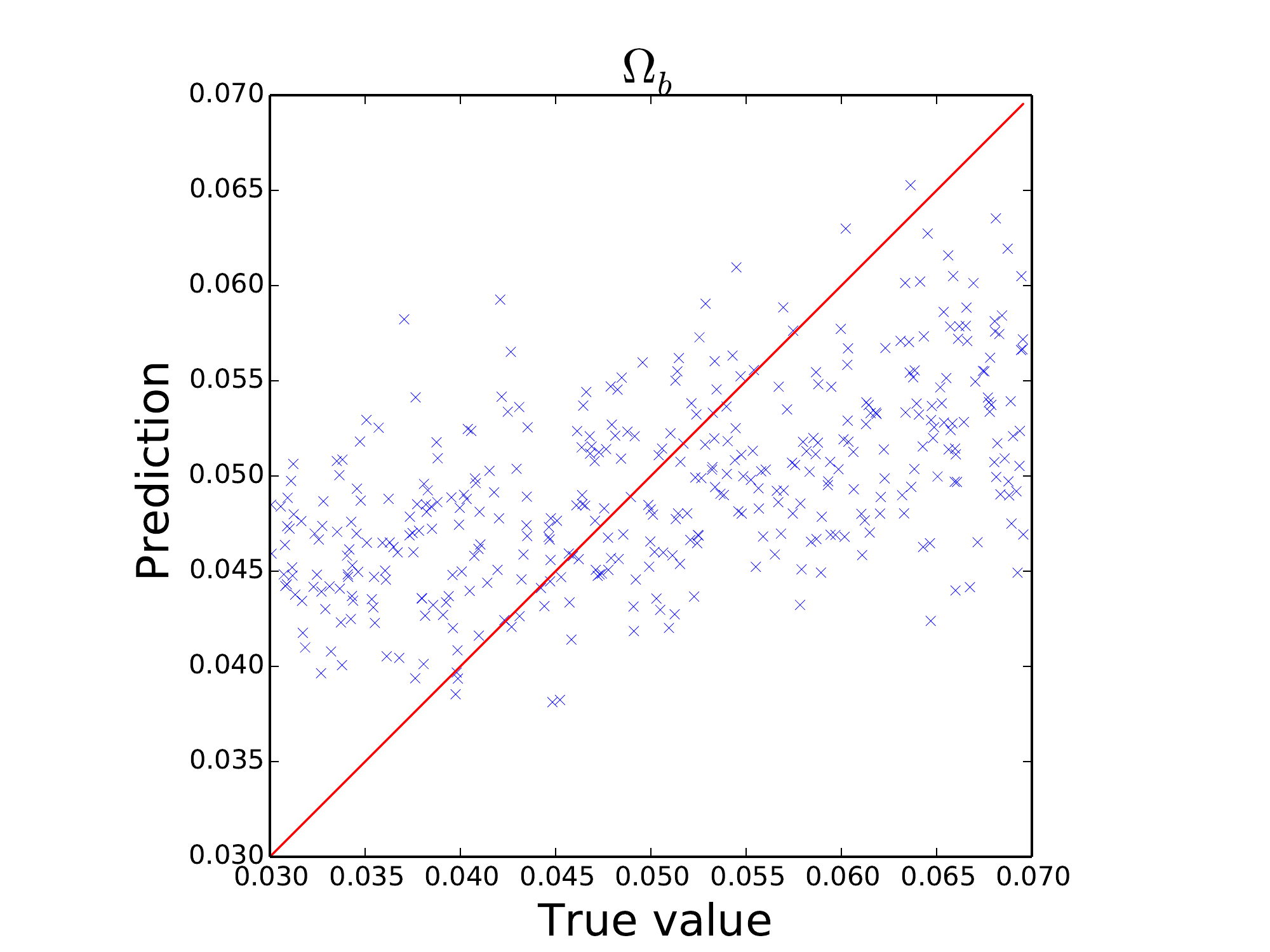} 
\includegraphics[width=0.49\linewidth]{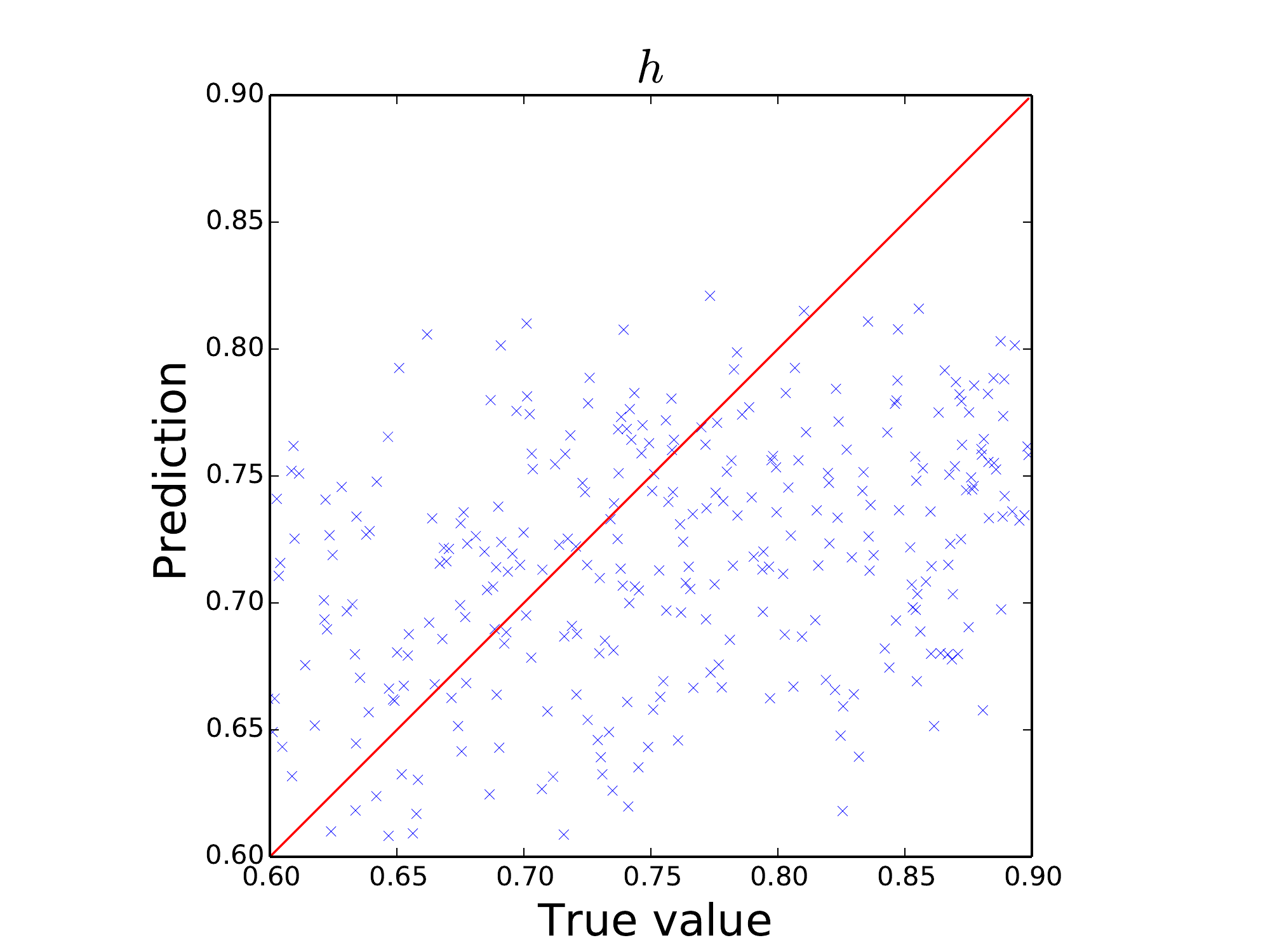} 
\includegraphics[width=0.49\linewidth]{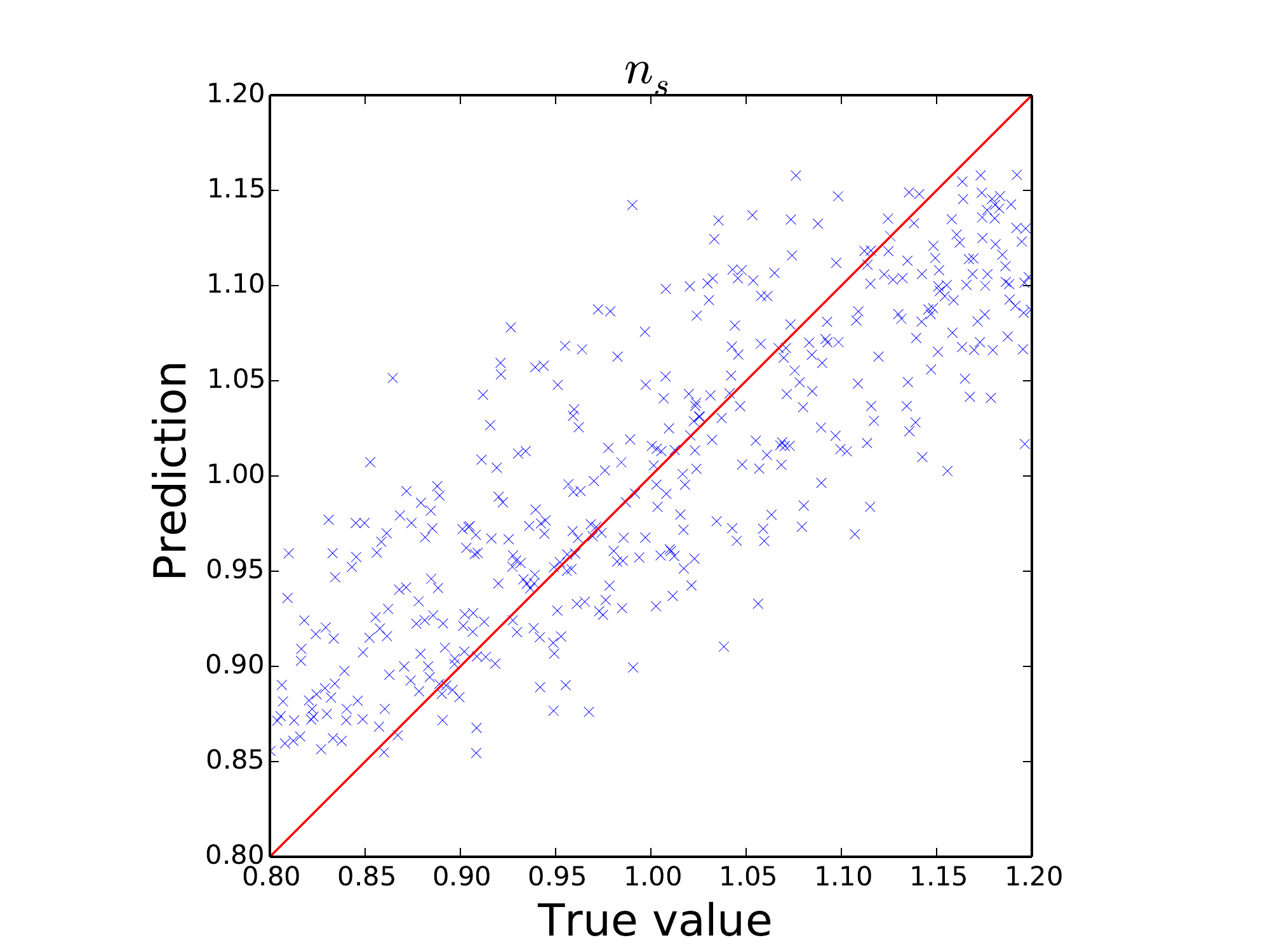} 
\includegraphics[width=0.49\linewidth]{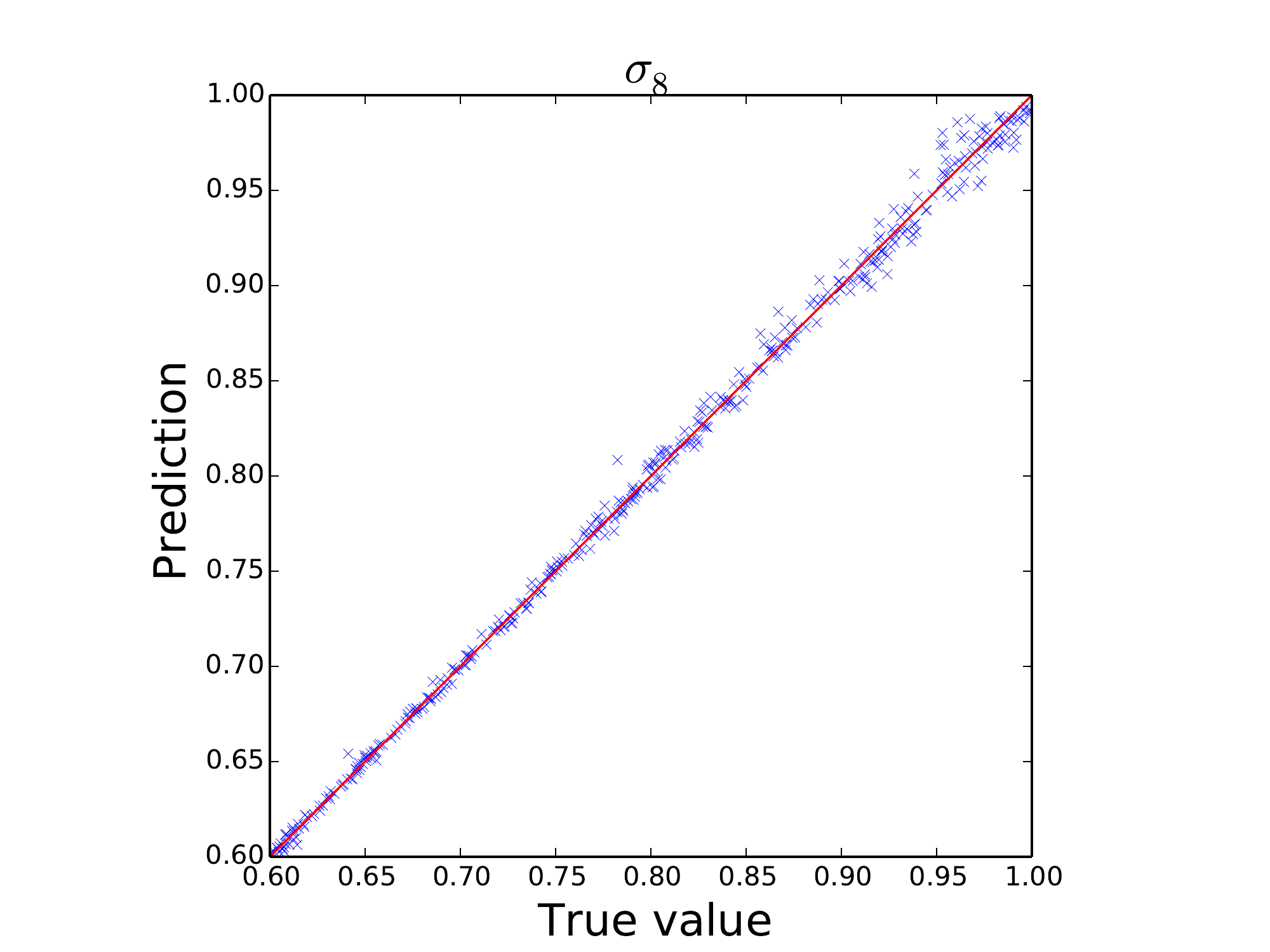} 
\caption{From top to bottom and left to right: Predictions for $\Omega_m$, $\Omega_b$, $h$, $n_s$ and $\sigma_8$ from the non-linear power spectrum extracted from high-resolution simulations using random forest regressors.}
\label{fig:ps5_rf}
\end{figure}

In the case of the random forest, the results that we have obtained confirm our findings from the simulations (Fig. \ref{fig:ps5_rf}). The RSEs obtained are at 0.09 for $\Omega_m$, 0.70 for $\Omega_m$, 0.70 for $h$, 0.33 for $n_s$ and 0.0025 for $\sigma_8$.     

\begin{figure}[!h]
\centering
\includegraphics[width=0.4\linewidth, trim=34cm 14cm 14cm 0, clip]{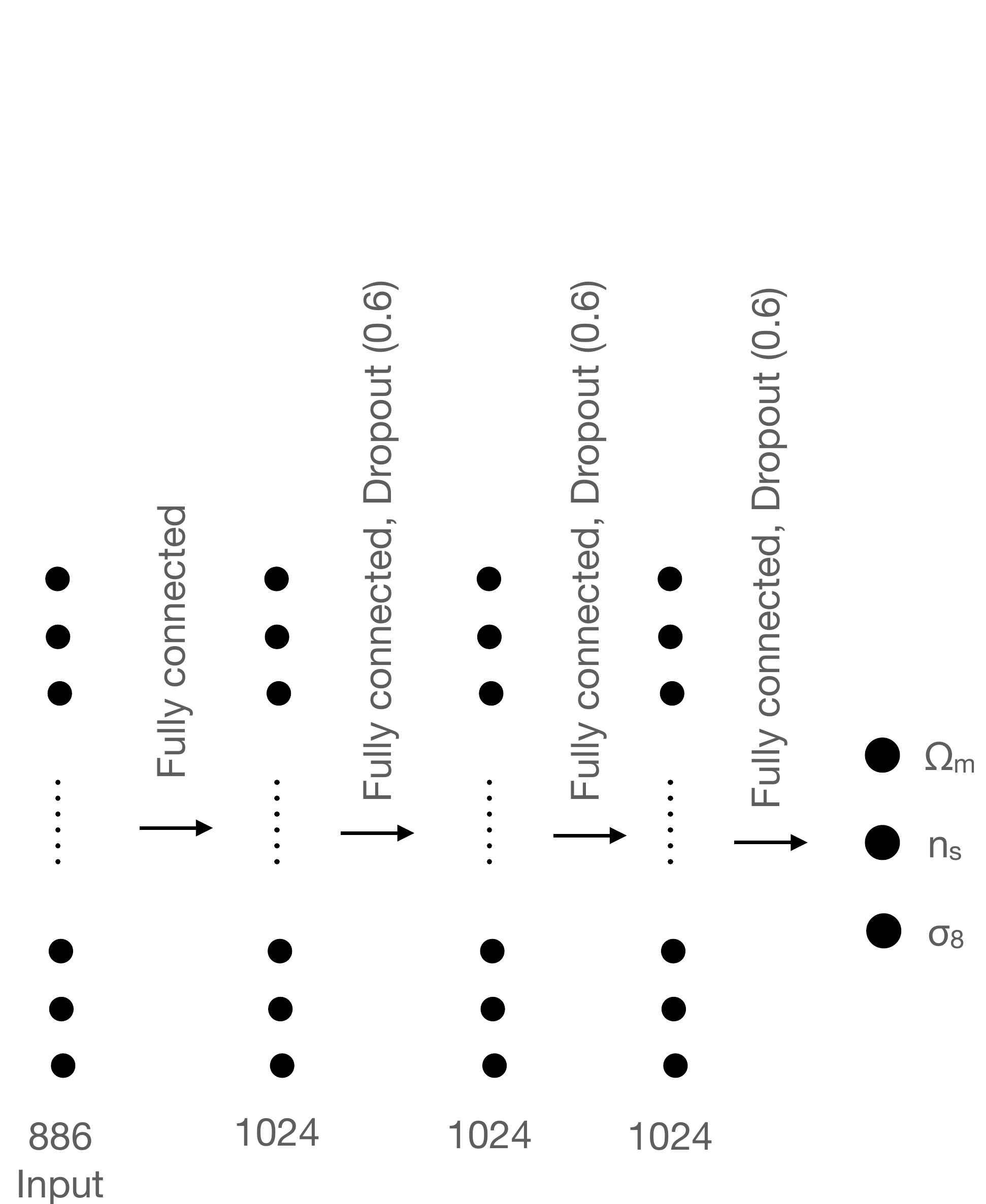}
\caption{Architecture of the deep neural network used to extract cosmological parameters from the power spectrum. Starting with the values of the spectra in 886 neurons, the network consists of three fully connected dense layers of 1024 neurons, followed Dropout layers where 60\% of the data is omitted. Finally, the network ends with the output layer of three neurons corresponding to $\Omega_m$, $n_s$ and $\sigma_8$.}
\label{fig:dnn}
\end{figure}

\begin{figure}[!h]
\centering
\includegraphics[width=0.49\linewidth]{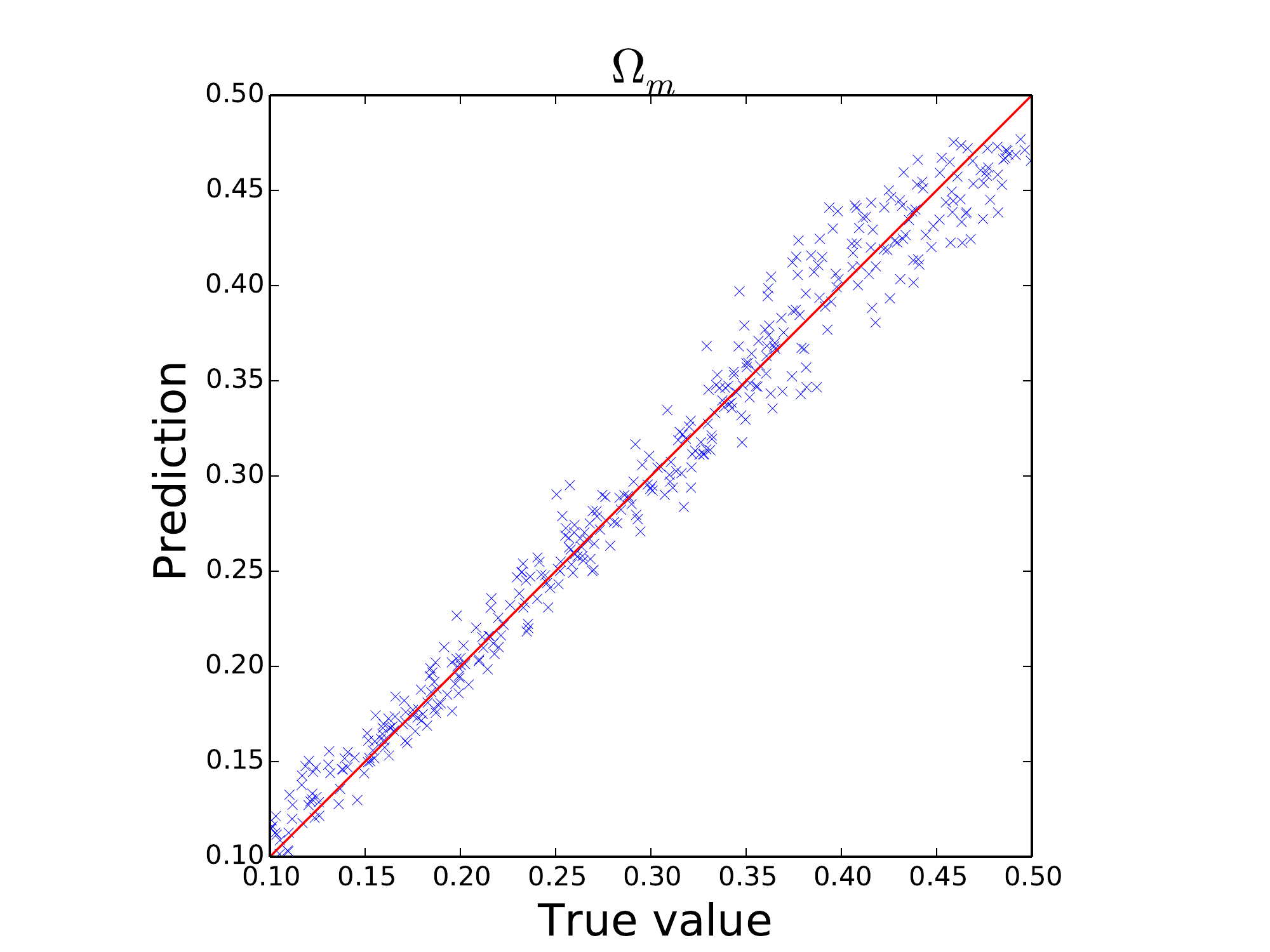} 
\includegraphics[width=0.49\linewidth]{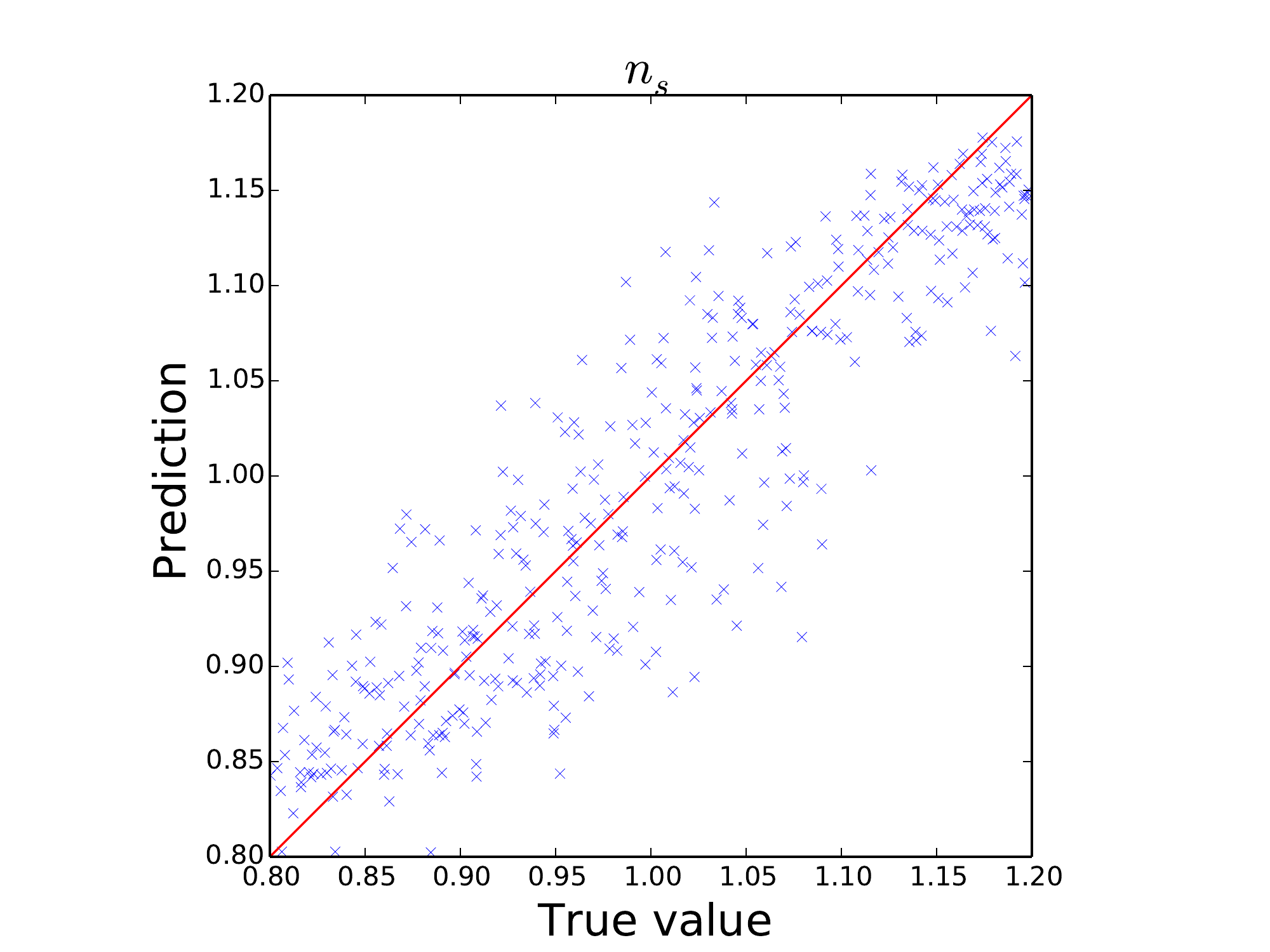} 
\includegraphics[width=0.49\linewidth]{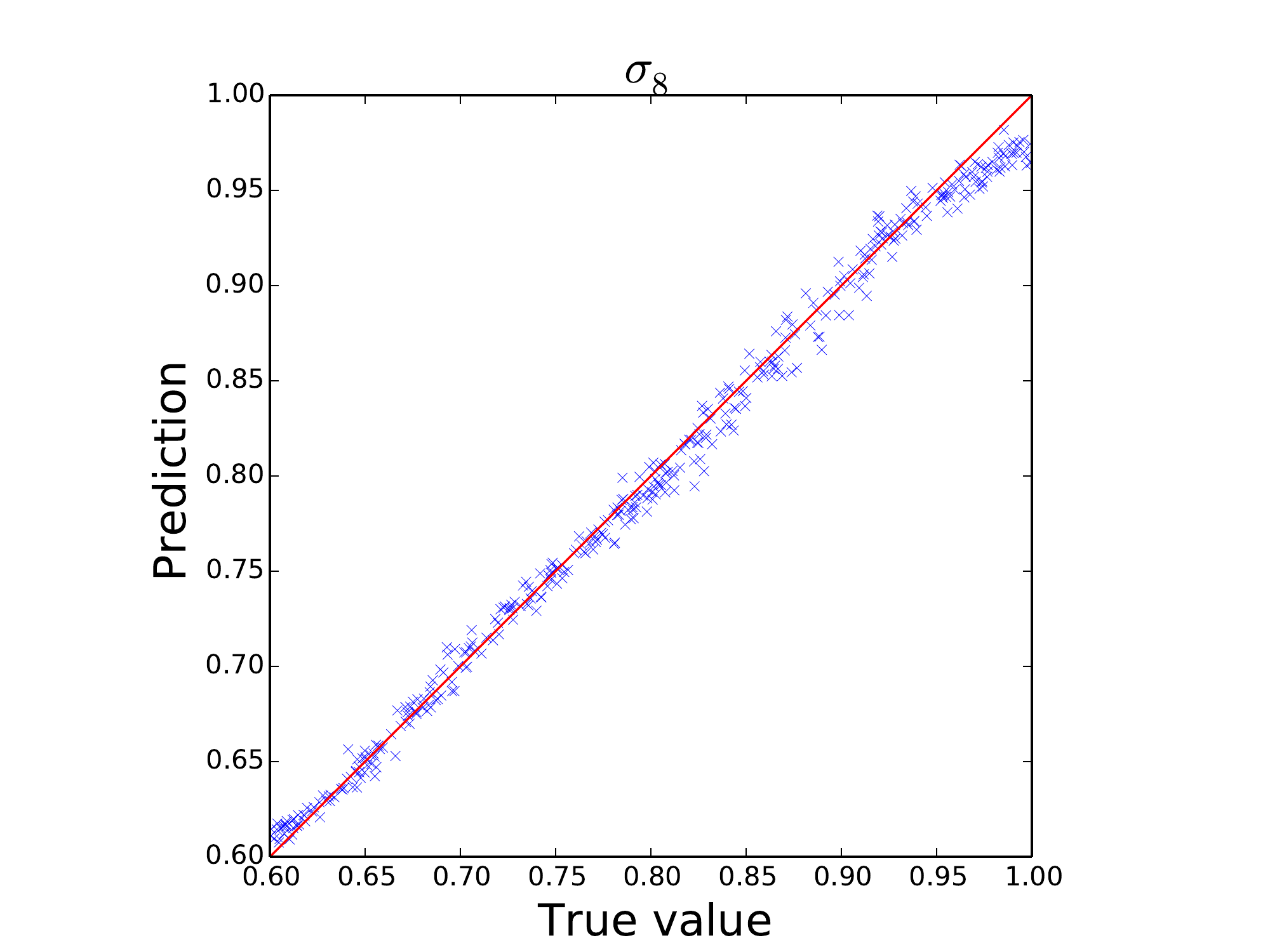} 
\caption{From top to bottom and left to right: Predictions for $\Omega_m$,  $n_s$ and $\sigma_8$ from the non-linear power spectrum extracted from high-resolution simulations using a deep neural network.}
\label{fig:ps3_dnn}
\end{figure}

As the plots in Fig. \ref{fig:ps5_rf} show that $\Omega_b$ and $h$ are not well determined, we skip these parameters when running our deep neural network. On the other hand, as the error for $n_s$ is significantly reduced, when we employ a deep neural network for the power spectrum, we include $\Omega_m$, $n_s$ and $\sigma_8$. We consider a simple neural network, consisting of three 1024-neurons hidden layers, each followed  by a Dropout layer \cite{JMLR:v15:srivastava14a}, with rates of 0.6 (Fig. \ref{fig:dnn}). 
The input layer has a size of 886 (the number of power spectrum bins), and the output layer has size 3. We are employing an Adam optimizer with a learning rate of $5 \times 10^{-6}$ \cite{Kingma:2014vow} and default first and second moment exponential decay rates (0.9 and 0.999). The results that we have obtained are presented in Fig. \ref{fig:ps3_dnn}, where the model has been run for 900 epochs, and the RSEs are 0.022 for $\Omega_m$, 0.17 for $n_s$ and 0.0057 for $\sigma_8$.

\section{Discussion}
\label{sec:disc}

In Table \ref{tab:allres} we quantify the goodness-of-fit and we present the relative squared errors obtained from the three dimensional density field (via a convolutional neural network) and from the power spectrum (using the random forest regressor and the deep neural network). These results show a significantly better accuracy of the estimation of the parameters involved when using the power spectrum, compared to using the convolutional neural network on the 3D density field. We believe that the lower resolution of the simulations are the cause for the weaker accuracy in determining parameters from the simulations with respect to the power spectrum. In the case of the power spectrum, the inability of the network to extract $\Omega_b$ and $h$ is caused by the degeneracies between them \cite{Brieden:2021edu}. As $\sigma_8$ modifies the global amplitude of the power spectrum, the accurate determination of this parameter was expected. For $n_s$, which only changes the tilt of the linear power spectrum, we have obtained a much weaker determination, likely due to non-linearities and degeneracies with the other parameters. Finally, increasing $\Omega_m$ shifts power into smaller scales, hence providing an observable effect. 

\begin{table}[!h]
\begin{center}
\begin{tabular}{c|P{28mm}|P{28mm}|P{28mm}|P{28mm}}\hline
    Parameter & Simulations (5 parameters) & Simulations  (2 parameters) & Power spectrum (RF) & Power spectrum (DNN) \\ \hline\hline
    $\Omega_m$ &0.22 & 0.23 & 0.09 & 0.022  \\
    \hline
    $\Omega_b$ &1.19& -- & 0.70 & --  \\
    \hline
    $h$ &1.35& -- & 0.70 & --  \\
    \hline
    $n_s$ &1.27& -- & 0.33 & 0.17  \\
    \hline
    $\sigma_8$ & 0.025 & 0.015 & 0.0025 & 0.0067  \\
    \hline
\end{tabular}
\caption{Relative squared errors on the cosmological parameters obtained from the convolutional neural network using the simulations (left), and from the power spectrum using a random forest regressor (middle) and a deep neural network (right).}
\label{tab:allres}
\end{center}
\end{table}

\section{Conclusions and future directions}
\label{sec:conc}
In this work, we have shown how information from numerical simulations can be extracted either directly, or after processing, using correlation functions such as the matter power spectrum.

Our results show that, among the parameters that are varied in the \textsc{Quijote} simulations, $\sigma_8$ can be extracted with exquisite accuracy, from both simulations (through the 3D density field) and the power spectrum.
The matter fraction $\Omega_m$ can also be extracted from both the simulations and the power spectrum, although less accurately. We also show that the spectral index can be determined from the power spectrum using a deep neural network. Of course, the information stored in the non-linear power spectrum also appears in the density field, and therefore we conclude that the low resolution ($64^3$) does not capture all the relevant information from the simulation. As the training time and the memory requirements of increasing the input size from $64^3$ points to $128^3$ or $256^3$ for each example and using them directly would be significantly increased and a very different network configuration is likely to be required, we leave such a study for future work, where we plan to investigate the optimal methods of extracting the cosmological parameters using convolutional neural networks. 

These type of methods could be eventually used to determine the parameters from measured galaxy patterns on the sky. The methods would require further refining as in this work we have only focused on the matter distribution. A study of the bias between the dark matter and galaxy distributions would also be required.

\acknowledgments
AL acknowledges funding by the LabEx ENS-ICFP: ANR-10-LABX-0010/ANR-10-IDEX-0001-02 PSL*.

\bibliography{Bibliography}{}

\end{document}